\documentstyle[12pt,epsfig]{article}

\newlength{\dinwidth}
\newlength{\dinmargin}
\begin{document}
%
%
%
%
%
%
\begin{titlepage}
\title{\bf Measurement of Elastic $\phi$ Photoproduction at HERA}

\author{\large{\rm ZEUS Collaboration} }

\date{}
\maketitle
\vspace{5cm}

\begin{abstract}

\noindent
The production of $\phi$ mesons in the reaction $e^{+}p \rightarrow e^{+}
\phi p$ ($\phi \rightarrow K^{+}K^{-}$) at a median $Q^{2}$ of $10^{-4} \
\rm{GeV^2}$ has been studied with the ZEUS detector at HERA.  The
differential $\phi$ photoproduction cross section $d\sigma/dt$ has an
exponential shape and has been determined in the kinematic range
$0.1<|t|<0.5 \ \rm{GeV^2}$ and $60 < W < 80 \ \rm{GeV}$. An integrated
cross section of $\sigma_{\gamma p \rightarrow \phi p} = 0.96 \pm
0.19^{+0.21}_{-0.18}$ $\rm{\mu b}$ has been obtained by extrapolating to
{\it t} = 0. When compared to lower energy data, the results show a weak
energy dependence of both $\sigma_{\gamma p \rightarrow \phi p}$ and the
slope of the $t$ distribution. The $\phi$ decay angular distributions are
consistent with $s$-channel helicity conservation. From lower energies to
HERA energies, the features of $\phi$ photoproduction are compatible with
those of a soft diffractive process. 

\end{abstract}
\vspace{-19cm}
\begin{flushleft}
\tt DESY 96-002 \\
January 1996 \\
\end{flushleft}

\setcounter{page}{0}
\thispagestyle{empty}
\eject

\end{titlepage} 

\def\3{\ss}
\textwidth 15.5cm
\parindent 0cm

\footnotesize
\renewcommand{\thepage}{\Roman{page}}
\begin{center}
\begin{large}
The ZEUS Collaboration
\end{large}
\end{center}
M.~Derrick, D.~Krakauer, S.~Magill, D.~Mikunas, B.~Musgrave,
J.R.~Okrasinski, J.~Repond, R.~Stanek, R.L.~Talaga, H.~Zhang \\
{\it Argonne National Laboratory, Argonne, IL, USA}~$^{p}$
% ...
\\[6pt]
\clearpage
\hspace*{1mm}
% ...
$^{ 2}$ now at Univ. of Salerno and INFN Napoli, Italy  \\
\hspace*{1mm}
% ...

\begin{tabular}{lp{15cm}}
$^{a}$ & supported by the Natural Sciences and Engineering Research
         Council of Canada (NSERC) \\
$^{b}$ & supported by the FCAR of Qu\'ebec, Canada\\
$^{c}$ & supported by the German Federal Ministry for Education and
         Science, Research and Technology (BMBF), under contract
         numbers 056BN19I, 056FR19P, 056HH19I, 056HH29I, 056SI79I\\
$^{d}$ & supported by the MINERVA Gesellschaft f\"ur Forschung GmbH,
         and by the Israel Academy of Science \\
$^{e}$ & supported by the German Israeli Foundation, and
         by the Israel Academy of Science \\
$^{f}$ & supported by the Italian National Institute for Nuclear Physics
         (INFN) \\
$^{g}$ & supported by the Japanese Ministry of Education, Science and
         Culture (the Monbusho)
         and its grants for Scientific Research\\
$^{h}$ & supported by the Korean Ministry of Education and Korea Science
         and Engineering Foundation \\
$^{i}$ & supported by the Netherlands Foundation for Research on Matter
         (FOM)\\
$^{j}$ & supported by the Polish State Committee for Scientific
         Research, grants No.~115/E-343/SPUB/P03/109/95, 2P03B 244
         08p02, p03, p04 and p05, and the Foundation for Polish-German
         Collaboration (proj. No. 506/92) \\
$^{k}$ & supported by the Polish State Committee for Scientific
         Research (grant No. 2 P03B 083 08) \\
$^{l}$ & partially supported by the German Federal Ministry for
         Education and Science, Research and Technology (BMBF) \\
$^{m}$ & supported by the German Federal Ministry for Education and
         Science, Research and Technology (BMBF), and the Fund of
         Fundamental Research of Russian Ministry of Science and
         Education and by INTAS-Grant No. 93-63 \\
$^{n}$ & supported by the Spanish Ministry of Education and Science
         through funds provided by CICYT \\
$^{o}$ & supported by the Particle Physics and Astronomy Research
         Council \\
$^{p}$ & supported by the US Department of Energy \\
$^{q}$ & supported by the US National Science Foundation
\end{tabular}

\clearpage

\pagenumbering{arabic}
\setcounter{page}{1}
\normalsize

%
%
%
%
%
%
%
\section{Introduction}
Elastic photoproduction of $\phi$ mesons, $\gamma p \rightarrow \phi p$,
has previously been studied at photon-proton centre-of-mass 
energies up to $W \approx 17$ GeV 
in fixed target experiments~\cite{bauer,busen,egloff}.
At these energies
the reaction $\gamma p \rightarrow \phi p$ has the characteristics
of a soft diffractive process: a cross section rising weakly with the 
centre-of-mass energy, a steep forward diffractive peak in the
$t$ distribution, where $t$ is the 
four-momentum transfer squared at the proton vertex,
and $s$-channel helicity conservation.
\par\medskip\noindent
At the energies of the previous experiments,
elastic vector meson photoproduction 
is well described as a soft diffractive process
in the framework of the Regge theory~\cite{regge}
and of the Vector Dominance Model (VDM)~\cite{vdm}.
In this approach, this reaction 
is assumed to proceed at high energy through the exchange of a ``soft''
pomeron trajectory~\cite{dl} with an 
effective intercept of $1+ \epsilon = 1.08$ and a slope of 
$\alpha' = 0.25$ $\rm{GeV^{-2}}$.
Recent experimental results~\cite{rho} extend the validity of
this approach to the high energies of HERA for elastic $\rho^{0}$
photoproduction. 
This approach also provides predictions~\cite{dl} for the total
photoproduction cross section consistent with the measurements made
at HERA~\cite{sphp}.
In contrast, the predictions of Regge
theory fail to describe the recently measured rapidly rising
cross sections at HERA for elastic $J/\psi$ 
photoproduction~\cite{psi} and for exclusive 
$\rho^0$ production
($\gamma^{\star} p \rightarrow \rho^{0} p$)
in deep inelastic scattering (DIS)~\cite{rhodis}. The
measurements for the latter two processes are consistent with perturbative QCD
calculations~\cite{ryskin,brodski}. In these
calculations the scale of the process is given by the virtuality $Q^2$ of the
exchanged photon for exclusive DIS $\rho^0$ 
production or depends on the mass of
the vector meson for elastic $J/\psi$ photoproduction. 
Perturbative QCD calculations for the proton structure function $F_2$
are consistent with the data~\cite{f2} at HERA energies for a scale
as small as $Q^2 = 1.5$ $\rm{GeV^2}$. If the scale of elastic vector
meson photoproduction is given by the mass of the vector meson, 
the scale of elastic $\phi$ photoproduction is between that
of elastic $\rho^0$ and $J/\psi$ photoproduction and between
that of elastic $\rho^0$ photoproduction and exclusive DIS
$\rho^0$ production.
It is therefore of interest to measure elastic $\phi$ photoproduction
and to see whether the scale
of the process
is large enough to cause a
deviation from the behavior of a soft diffractive process.
\par\medskip\noindent
The expectations of Regge theory and VDM may also be confronted
by a measurement of 
elastic $\phi$ photoproduction at HERA energies. In
the additive quark model~\cite{aqm} and VDM, 
the reaction $\gamma p \rightarrow \phi p$
can proceed only by pomeron exchange~\cite{freund}, and is thus
a particularly clean example of a diffractive reaction.
\par\medskip\noindent
This paper reports a measurement with the ZEUS detector at HERA
of high energy production of $\phi$
mesons in the reaction $e^{+}p\rightarrow e^{+}\phi p$ 
using events with $Q^2 < 4~{\rm GeV^2}$
in which neither the scattered proton nor the scattered positron
was observed in the detector.
The $\phi$ was observed,
via its decay into $K^+K^-$, in the kinematic range $60 < W < 80$ GeV
and $0.1 < p_T^2 < 0.5$ $\rm{GeV^2}$, 
where $p_T$ is the transverse momentum of the $\phi$ 
with respect to the beam axis. 
%
%
%
%
%
%
\section{Kinematics}
Figure 1 shows a schematic diagram for the reaction
\[
e^{+}(k)p(P) \rightarrow e^{+}(k')\phi(V)p(P'),
\]
where each quantity in parentheses is the four-momentum of the particle.
\par\medskip\noindent
The kinematics of the inclusive scattering of 
unpolarised positrons and protons is
described by the positron-proton centre-of-mass energy
and any two of the following variables:
\begin{itemize}
\item $Q^2=-q^2=-(k-k')^2$, the negative of the 
four-momentum squared of the exchanged photon;
\item $y=(q\cdot P)/(k\cdot P)$, the fraction 
of the positron energy transferred
to the hadronic final state in the rest frame of the initial state proton;
\item
$W^2 = (q+P)^2= -Q^2+2y(k\cdot P)+M^2_p$, the centre-of-mass energy squared
of the photon-proton system, where 
$M_p$ is the proton mass. 
\end{itemize}
\par\medskip\noindent
For the description of the exclusive reaction
$e^{+}p \rightarrow e^{+}\phi p$ ($\phi \rightarrow K^{+}K^{-}$)
the following additional variables are required:
\begin{itemize}
\item 
$t = (P-P')^2$, 
the four-momentum transfer squared at the proton vertex;
\item
the angle between the $\phi$ production plane and the positron 
scattering plane;
\item
the polar and azimuthal angles of the decay kaons in the $\phi$ rest frame.
\end{itemize}
In the present analysis, events were selected in which the final state positron
was scattered at an angle too small to be detected in the 
main ZEUS calorimeter. Thus
the angle between the $\phi$ production plane and the positron 
scattering plane was not measured.
In such untagged photoproduction events
the $Q^2$ value ranges from the kinematic minimum 
$Q^2_{min} = M^2_e y^2/(1-y) 
\approx 10^{-9} \ \rm{GeV^2}$, where $M_e$ is the electron mass, 
to the detector limit 
$Q^2_{max} \approx 4 \ \rm{GeV^2}$,
with a median $Q^2$ of approximately $10^{-4} \ \rm{GeV^2}$.
Because of this small $Q^2$ value,
the photon-proton centre-of-mass energy can be expressed as:
$$
W^2 \simeq 2 (E_{\phi} - p_{Z \phi}) E_p,
$$
where $E_p$, $E_{\phi}$ are the energies of the 
incoming proton and the produced
$\phi$ meson and $p_{Z \phi}$ 
\footnote{Throughout this paper use is made of the standard ZEUS right-handed
coordinate system in which the positive $Z$-axis points in the direction
of flight of the protons (referred to as the 
forward direction) and the $X$-axis
is horizontal, pointing towards the center of HERA. The nominal interaction
point is at $X=Y=Z=0$.}
is the longitudinal momentum of the $\phi$ meson.
Similarly, the four-momentum transfer squared, 
{\it t}, at the proton vertex for $Q^2 = Q^2_{min}$
is given by:
$$ {\it t} = (q-V)^2 \simeq -p^2_T \ .$$
Non-zero values of $Q^2$ cause $t$ to differ from $-p^2_T$
by less than $Q^2$,
as described elsewhere~\cite{rho}.
%
%
%
%
%
%
\subsection{{\boldmath $\phi$} photoproduction}
The $\gamma p$ cross section is related to the $e^{+} p$
cross section by:
\[
\frac{d^{2}\sigma^{ep}}{dydQ^{2}}=\left.\frac{\alpha}{2\pi}\frac{1}{Q^{2}}\right[ \
\left( \frac{1+(1-y)^{2}}{y}-\frac{2(1-y)}{y}\frac{Q^{2}_{min}}{Q^{2}}\right) \
\sigma_{T}^{\gamma^{\star}p}(W,Q^{2})+ 
\]
\[
\left.\frac{2(1-y)}{y}\sigma_{L}^{\gamma^{\star}p}(W,Q^{2})\right] 
\]
where $\sigma_{T}^{\gamma^{\star}p}$ and $\sigma_{L}^{\gamma^{\star}p}$
are the $\gamma p$ cross sections for transversely and longitudinally 
polarized photons, respectively.
\par\medskip\noindent
Using the VDM predictions~\cite{vdm}:
\[ 
\frac{\sigma_{L}^{\gamma^{\star}p}(W,Q^{2})}{ 
\sigma_{T}^{\gamma^{\star}p}(W,Q^{2})} = \xi \frac{Q^{2}}{M_{\phi}^{2}}
\]
\[
\sigma_{T}^{\gamma^{\star}p}(W,Q^{2})=
\frac{\sigma_{T}^{\gamma p}(W,0)}{(1+Q^{2}/M_{\phi}^{2})^{2}} \equiv
\frac{\sigma^{\gamma p}(W)}{(1+Q^{2}/M_{\phi}^{2})^{2}}
\]
where $M_{\phi}$ is the $\phi$ meson mass, 
and using $\xi =1$~\cite{xi} yields
\[
\frac{d^{2}\sigma^{ep}}{dydQ^{2}} = F(y,Q^{2})\sigma^{\gamma p}(W)
\]
where the function:
\[
F(y,Q^{2})=\left.\frac{\alpha}{2\pi}\frac{1}{Q^{2}}\right[ \
\left(\frac{1+(1-y)^{2}}{y}-\frac{2(1-y)}{y}(\frac{Q^{2}_{min}}{Q^{2}}-
\frac{Q^{2}}{M_{\phi}^{2}})\right)\left.\frac{1}{(1+Q^{2}/M_{\phi}^{2})^{2}}\right]
\]
is the effective photon flux.
\par\medskip\noindent
Assuming no strong {\it W} dependence,
the $\gamma p$ cross section 
at the average {\it W} measured in the experiment
is obtained as the ratio of the
corresponding $e^+p$ cross section, integrated over the
{\it y} and $Q^2$ ranges covered by the measurement, and the photon
flux factor $F(y,Q^{2})$ integrated over the same {\it y} and $Q^2$
ranges.
%
%
%
%
%
%
%
\section{Experimental setup}
\subsection{HERA}
During 1994 HERA operated with a proton beam energy of 820 GeV and a positron
beam energy of 27.5 GeV.
In the positron and the proton beams
153 colliding bunches
were stored 
together with an additional 17 proton and 15 positron unpaired
bunches. These additional bunches were
used for background studies. The time between bunch crossings
was 96 ns.
The typical instantaneous luminosity was $1.5 \cdot 10^{30}$
$\rm{cm^{-2}s^{-1}}$.
\subsection{The ZEUS detector}
A detailed description of the ZEUS detector can be found elsewhere~\cite{zeus}.
The main components used in this analysis are outlined below.
\par\medskip\noindent
Charged particle~ momenta~ are~ reconstructed~ by the vertex 
detector (VXD)~\cite{vxd}, the central
tracking detector (CTD)~\cite{ctd} and 
the rear tracking detector (RTD)~\cite{zeus}. 
The VXD and the CTD are  cylindrical drift chambers
placed in a magnetic field of 1.43 T
produced by a thin superconducting coil. The vertex detector surrounds
the beam pipe and consists of 120 radial cells, each with 12 sense
wires.
The CTD surrounds the vertex detector and consists
of 72 cylindrical layers, organized in 9 superlayers
covering the polar angular region $15^o < \theta < 164^o$.
The RTD is a planar drift chamber located at the rear of the CTD
covering the polar angular region $162^o < \theta < 170^o$.
Using the information from the VXD, the CTD and the RTD
for the two-track events of this analysis,
the primary event vertex
was reconstructed 
with a
resolution of 1.4 cm in Z and 0.2 cm in the transverse plane.
\par\medskip\noindent
The high resolution uranium-scintillator calorimeter CAL~\cite{cal} is divided into
three parts, the forward calorimeter (FCAL), the barrel calorimeter (BCAL)
and the rear calorimeter (RCAL), which 
cover polar angles from  $2.6^o$ to $36.7^o$,
$36.7^o$ to $129.1^o$, and $129.1^o$
to $176.2^o$, respectively. 
Each part consists of towers which are longitudinally subdivided
into electromagnetic (EMC) and hadronic (HAC) readout cells.
The transverse sizes are typically $5\times20$ $\rm{cm^2}$ for the EMC cells
($10\times20$ $\rm{cm^2}$ in RCAL) and $20\times20$ $\rm{cm^2}$ for the HAC cells.
From test beam data, energy resolutions with $E$ in GeV of
$\sigma_{E}/E = 0.18/\sqrt{E}$ for electrons and $\sigma_{E}/E = 0.35/\sqrt{E}$
for hadrons have been obtained.
\par\medskip\noindent
Proton-gas events occuring upstream of the nominal $e^{+} p$
interaction point are out of time with respect to the $e^{+} p$
interactions and may thus be rejected 
by timing measurement made
by the scintillation
counter arrays Veto Wall, C5 and
SRTD respectively situated along the beam line at $Z = -730$ cm,
$Z = -315$ cm and $Z = -150$ cm.
\par\medskip\noindent
The luminosity is determined~\cite{lumi} from the rate of the 
Bethe-Heitler process $e^{+} p \rightarrow e^{+} \gamma p$
where the photon is measured by the LUMI calorimeter located
in the HERA tunnel in the direction
of the positron beam. 
%
%
%
%
%
%
\subsection{Trigger}
ZEUS has a three level trigger system.
The data used in this analysis were taken with the ``untagged
vector meson photoproduction trigger'' described in this section.
The photoproduction events are ``untagged'' since the scattered
positron escapes undetected through the beam pipe hole in the RCAL.
\par\medskip\noindent
The first level trigger (FLT) required:
\begin{itemize}
\item 
a minimum energy deposit of 464 MeV in the electromagnetic section 
of the RCAL, excluding the towers immediately surrounding the beam pipe;
\item 
at least one track candidate in the CTD;
\item 
less than 1250 MeV deposited in the FCAL towers surrounding
the beam pipe.
\item
the time of any energy deposited in the Veto Wall, the C5
and the SRTD to be consistent with an $e^{+} p$ interaction
and not with a proton-gas event.
\end{itemize}
\par\medskip\noindent
The second level trigger (SLT) rejected background events exploiting the
excellent time resolution of the calorimeter.
\par\medskip\noindent
The third level trigger (TLT) used information from the CTD
to select events with a reconstructed  vertex,
at most 4 reconstructed tracks and an invariant mass 
less than 1.5 GeV for
at least one
two-track
combination assuming that the particles are pions.
The rate of the untagged vector meson photoproduction trigger leaving the TLT 
was about 2 Hz. Because of this high rate 
the trigger was prescaled.
The recorded data collected during 1994 from this trigger 
correspond to an effective integrated 
luminosity of $887 \pm 31$ $\rm{nb^{-1}}$.
%
%
%
%
%
%
%
\section{Event selection}
The following offline cuts were applied to select the reaction
$\gamma p \rightarrow \phi(\rightarrow K^+K^-) p$:
\begin{itemize}
\item
exactly two oppositely charged tracks associated 
with a reconstructed vertex;
\item
each track within the pseudorapidity\footnote{The pseudorapidity $\eta$
is defined as $\rm{\eta = -ln[tan(\frac{\theta}{2})]}$.}
range $\left| \eta \right| < 2.0$
and with a transverse momentum
greater than 150 MeV. These cuts select the high
efficiency and well understood region  of the tracking detector;
\item
the Z coordinate of the vertex within $\pm30$ cm of the nominal
interaction point;
\item
in BCAL and RCAL, not more than 200 MeV in any 
EMC (HAC) calorimeter cell which is more than
30 cm (50 cm) away from the extrapolated impact position of
either of the two tracks.
This cut rejects events with additional neutral particles;
\item
energy deposit in FCAL less than 0.8 GeV. This cut reduces the contamination
from diffractive proton dissociation, $\gamma p \rightarrow \phi X$.
\end{itemize}
\par\medskip\noindent
Since the detector geometry and the trigger limit
the observable kinematic range for the reaction $\gamma p \rightarrow \phi p$,
the selected events were restricted to the region:
\[ 
60 < W < 80 \: \rm{GeV}
\]
\[
0.1 < p_T^2 < 0.5 \: \rm{GeV^2},
\] 
where the acceptance is well understood and the background
contamination due to proton dissociation 
is relatively small.
%
%
%
%
%
%
%
\section{Monte Carlo simulation and acceptance calculation}
The acceptance for untagged elastic $\phi$ photoproduction was 
calculated by Monte Carlo methods. The reaction 
$e^{+}p \rightarrow e^{+}\phi p$ was simulated using two different event
generators.
The first one, DIPSI, is based on a model by Ryskin~\cite{ryskin}. It 
describes elastic vector meson production by the exchange of a pomeron
which interacts with 
the quark-antiquark pair into
which the incoming virtual photon fluctuates.
The second generator, JETPHI, uses a VDM approach and 
was written in the framework of the JETSET package~\cite{jetset}.
\par\medskip\noindent
Events were generated in the $W$ range from 50 to 90 GeV and the
$Q^2$ range between $Q^2_{min}$ and 4 $\rm{GeV^2}$.
The $\phi$ decay angular distributions in both programs were 
simulated assuming $s$-channel helicity conservation. 
To reproduce the $p_T^2$ distribution of the data,
the {\it t} dependence was taken to be of the form
$e^{-b|t|}$ with $b = 7$ $\rm{GeV^{-2}}$.
The input vertex distribution was simulated in accordance
with that measured using non-diffractive photoproduction
events.
\par\medskip\noindent
The generated events were processed
through the ZEUS detector and trigger simulation programs
as well as through the analysis chain. 
The same offline cuts were used for the Monte Carlo events and for the data.
The reconstructed $W$, $p_T^2$ and decay angular distributions
of the Monte Carlo sample agree well with those of the data.
\par\medskip\noindent
The acceptance as a function of $W$ and $p_{T}^2$
is shown in Fig. 2. 
The acceptance drops at low values of $W$ because
the decay kaons enter BCAL, not RCAL,
thus providing no trigger. At high $W$ as well as at low $p_{T}$ values
the acceptance decreases because the decay 
kaons emerge at a large polar angle and are not detected
by the CTD.
\par\medskip\noindent
The acceptance as a function of $M_{KK}$, the invariant mass of the
two decay kaons, 
is flat in the $\phi$
mass region.
%
%
%
%
%
%
\section{Results}
\subsection{Extraction of the {\boldmath $\phi$} signal}
After applying all selection cuts, the two particle invariant mass
was computed for each event, assuming that 
the two charged particles are kaons.
The invariant mass distribution is shown in Fig. 3.
The line is a fit to the function:
\[
dN/dM_{KK} = BW(M_{KK}) + BG(M_{KK}),
\]
where
the functions {\it BW} and {\it BG} describe the 
resonance shape and background,
respectively. 
The resonance shape was
described by a relativistic p-wave Breit-Wigner function convoluted
with a Gaussian. The width of the Breit-Wigner function was fixed
at the Particle Data Group value of $4.43 \pm 0.06$ MeV~\cite{pdg}.
The background, mainly due to the reaction
$\gamma p \rightarrow \rho^{0} p$,
was taken to be of the form:
\[
BG = {\alpha}(M_{KK}-2M_{K})^{\beta},
\]
where $M_{K}$ is the kaon mass. 
\par\medskip\noindent
The fit yields $566 \pm 31$ $\phi \rightarrow K^+K^-$ 
mesons after 
background subtraction. 
The $\phi$ mass obtained from the fit is
$M_\phi = 1.020 \pm 0.001$ $\rm{GeV}$,
with an r.m.s. of the Gaussian of 4 MeV, 
compatible with the experimental resolution.
The value of the free parameter 
$\beta$ obtained from the fit is $\beta = 0.97 \pm 0.09$.
%
%
%
%
%
%
%
%
\subsection{{\boldmath $\phi$} Events from background reactions}
The main source of background to the elastic $\phi$ 
reaction is
the process
$\gamma p \rightarrow \phi X$, where the proton diffractively
dissociates into a hadronic final state $X$ which is not detected
in the main calorimeter.
The background was
evaluated by comparing the number of $\phi$ events in the data, 
without the $E_{FCAL} < 0.8$ GeV cut,
to a Monte Carlo simulation using  the PYTHIA generator~\cite{pythia} 
of the diffractive proton dissociation 
reaction $\gamma p \rightarrow \phi X$.
The mass spectrum of the diffractive events was simulated according to 
a $d\sigma/dM_{X}^{2} \propto (1/M_{X})^{2.25}$ distribution~\cite{cdf}.
The $t$ dependence was parametrized with the form
$e^{-b|t|}$ with $b=4$ $\rm{GeV^{-2}}$.
The contamination
of the elastic $\phi$ photoproduction sample from  proton
dissociation\footnote{The proton diffractive dissociation
contamination in this measurement and in elastic $\rho^{0}$ photoproduction
~\cite{rho} of $(11 \pm 1(\rm{stat}) \pm 6(\rm{syst})) \%$ are 
compatible, taking into account
the different acceptance at low $p_T^2$
and the different $p_T^2$ regions of the two measurements.}
was estimated to be $(24 \pm 7(\rm{stat}) \pm 6(\rm{syst})) \%$ 
in the $p_T^2$ range between 0.1 and 0.5 $\rm{GeV^2}$.
The systematic error was estimated by varying the exponent
in the diffractive $M_{X}$ distribution between 2 and 2.5.
Similarly, the uncertainty due to varying the generated $t$ slope in the MC
sample from 4 $\rm{GeV^{-2}}$ to 3 $\rm{GeV^{-2}}$ 
has been included in the systematic error.
\par\medskip\noindent
For each bin in $p_T^2$, 
the number of events in the $\phi$ peak was corrected for
the background from the diffractive proton
dissociation reaction $\gamma p \rightarrow \phi X$. The $p^2_T$ behaviour
of this background was taken from the PYTHIA MC simulation.
\par\medskip\noindent
The background due to a $\phi$ meson produced in a beam-gas interaction
was estimated from the analysis of events
coming from unpaired bunches.
The values found are $1\%$ for positron- and $< 1\%$ for proton-gas interactions.
%
%
%
%
%
%
%
%
\subsection{Elastic {\boldmath $\phi$} photoproduction  
cross section}
The cross section for the reaction $\gamma p \rightarrow \phi p$
is given by
\[
\sigma_{\gamma p \rightarrow \phi p} = 
\frac{N_{\phi}}{L \cdot F \cdot BR}
\]
where $N_{\phi}$ is the number of
acceptance corrected $\phi$ events, 
{\it BR} is the branching ratio of 
the $\phi$ decay into $K^+$$K^-$ ($49.1\pm0.9\%$)~\cite{pdg},
{\it L} is the effective integrated luminosity and
$F = 0.025$
is the photon flux factor integrated over the phase space
determined by the selection cuts.
\par\medskip\noindent
An exponential fit to the acceptance corrected $p_T^2$ distribution
in the range $0.1 < p_T^2 < 0.5$ $\rm{GeV^2}$ gives the slope
value of $6.5 \pm 1.0(\rm{stat})$ $\rm{GeV^{-2}}$.
\par\medskip\noindent
The acceptance corrected {\it t} 
distribution was reconstructed from the measured $p^2_T$ spectrum 
using a bin-by-bin correction, given by the ratio of the generated 
{\it t} and  the reconstructed $p_{T}^2$ 
distributions in the MC sample. 
The acceptance corrected {\it t} distribution is shown in Fig. 4a. 
An  exponential fit of the form 
$d\sigma/d|t| = d\sigma/d|t|\vert_{t=0} \cdot e^{-b|t|}$ 
in the {\it $|t|$} range between 0.1 and 0.5 $\rm{GeV^2}$ yields:
\[
b = 7.3 \pm 1.0(\rm{stat}) \pm 0.8(\rm{syst})  \rm~{GeV^{-2}}
\]
\[
\frac{d\sigma_{\gamma p\rightarrow \phi p}}{d|t|}\Big\vert_{t=0} = 7.2 \pm 2.1(\rm{stat}) \pm 1.8(\rm{syst})~ \rm{\mu b}/ \rm{GeV^{2}}
\]
The systematic error of the slope parameter $b$ 
includes uncertainties from the acceptance
calculation ($6\%$) and the applied cuts ($9\%$). The systematic error
for $d\sigma/dt\vert_{t=0}$ is due to the acceptance
calculation ($14\%$), the applied cuts ($15\%$) and a normalization
uncertainty due to the calorimeter trigger ($12\%$),
the signal fitting procedure ($7\%$) and
the luminosity measurement ($3.5\%$). 
\par\medskip\noindent
Fig. 4b shows the value of the slope parameter $b$ measured by this
experiment together
with the results of lower energy 
photoproduction experiments~\cite{busen,slope}.
The ZEUS measurement, when compared to the fixed target measurements, 
shows a weak energy dependence of $b$,
as predicted
by Regge theory~\cite{regge}.
The slope parameter for elastic $\rho^0$ photoproduction~\cite{rho} is
$b_{\rho} = 9.9 \pm 1.2 \pm 1.4$ $\rm{GeV^{-2}}$ measured
in the range $|t| < 0.5$ $\rm{GeV^{2}}$ using a fit of the form
$e^{-b|t|+ct^{2}}$.
In the framework of geometric diffractive models, the slope obtained here 
for the $\phi$ meson, compared to that of the $\rho^0$, 
indicates that the radius of $\phi p$ interaction is smaller
than that of $\rho^0 p$.
%
%
%
%
%
\subsection{Total elastic {\boldmath $\phi$} photoproduction cross section} 
\par\medskip\noindent
The total elastic cross section
was obtained by extrapolating the differential cross section
to $t = 0$ assuming a simple 
exponential {\it t} dependence.
The resulting value of the cross section
is:

$$\sigma_{\gamma p \rightarrow \phi p} = 0.96 \pm 0.19(\rm{stat}) ^{+0.21}_{-0.18} (\rm{syst}) \ \rm{\mu b}$$ 
integrated over the range $|t| < 0.5$ $\rm{GeV^2}$ 
and at  an average W of 70 GeV.
\par\bigskip\noindent
The systematic error includes uncertainties from 
the acceptance calculation ($8\%$), the applied cuts ($8\%$)
and the normalization as described in section 6.3. 
The uncertainty from the proton dissociation background subtraction 
made in each bin of the $p_T^2$ distribution has been
included in the statistical error.
The uncertainty of the {\it t} extrapolation has been 
estimated by using a
fit of the form 
$e^{-b|t| +ct^{2}}$ with different values of {\it c}.
Changing the parameter {\it c} from 0 to 3 $\rm{GeV^{-4}}$
increases the cross section
by $10\%$. 
The range of the variation for the 
parameter $c$  was taken in accordance
with the results obtained in high energy $\rho^{0}$
photoproduction~\cite{rho}. 
This uncertainty has been included in the systematic error.
\par\medskip\noindent
The cross section for the process $\gamma p \rightarrow \phi p$ 
from this analysis is compared in Fig. 5  
to results at lower $\gamma p$ 
centre-of-mass energies~\cite{busen,lowe}.
The data show a weak energy dependence of the cross section 
from 2 GeV to 70 GeV, as predicted by Regge theory~\cite{regge}.
\par\medskip\noindent
The cross section ratio of elastic $\phi$
and $\rho^{0}$~\cite{rho} photoproduction at $W=70$ GeV
is $0.065 \pm 0.013(\rm{stat})$.
The same ratio measured
at  $W=17$ GeV~\cite{egloff}
is $0.076 \pm 0.010$. 
These results show that there is no significant energy
dependence of the $\phi / \rho^0$ photoproduction
cross section ratio in this $W$ range.
%
%
\subsection{Total {\boldmath $\phi p$} cross section}
Using VDM and the optical theorem the $\phi$ photoproduction cross
section at $t=0$ can be related to the total $\phi p$ cross
section by:
\[
\frac{d\sigma_{\gamma p \rightarrow \phi p}}{d|t|}\Big\vert_{t=0} = \frac{4\pi\alpha}{f_{\phi}^2}\cdot\frac{1+\eta^2}{16\pi}\cdot(\sigma_{tot}^{\phi p})^2
\]
where $\eta$ is the ratio of the real to the imaginary part
of the forward $\phi p$ elastic scattering amplitude,
$f_{\phi}^2/4\pi$ is the $\gamma \phi$ coupling constant and
$\alpha$ is the fine structure constant.
For pure pomeron exchange $\eta=0$. 
Taking $f_{\phi}^2/4\pi =18.4$ (see e.g.~\cite{bauer}, p. 393)
yields:
\[
\sigma_{tot}^{\phi p} = 19 \pm 7~\rm{mb}.
\]
\par\medskip\noindent
%
Using a parametrisation~\cite{dl} based on Regge theory, this result
is in agreement with the additive quark model which predicts
(see e.g.~\cite{feld})
$\sigma_{tot}^{\phi p} \simeq \sigma_{tot}^{K^{+}p} + \sigma_{tot}^{K^{-}n} - \sigma_{tot}^{\pi^{+}p} = 19.9$ $\rm{mb}$ at $W=70$ GeV.
The comparison of $\sigma_{tot}^{\phi p}$ to the total $\rho^0 p$
cross section, $\sigma_{tot}^{\rho^0 p} = 
28.0 \pm 1.2(\rm{stat}) \pm 2.8(\rm{syst})$ $\rm{mb}$~\cite{rho},
indicates that the $\phi p$ interaction radius
is smaller than that of $\rho^0 p$.
This is consistent with the comparison 
in section 6.3
of the $\phi$ and $\rho^0$ slopes.
%
%
%
\subsection{Decay angular distributions}
The $\phi$ decay angular distributions can be used to determine
elements of the $\phi$ spin-density matrix~\cite{angle}.
In the $s$-channel helicity frame the decay angle $\theta_{h}$
is defined  as the angle between the $K^+$ and the direction of the
recoil proton in the $\phi$ centre-of-mass frame, while the azimuthal
angle $\phi_{h}$ is the angle between the $\phi$ decay plane and
the $\gamma \phi$ plane in the $\gamma p$ centre-of-mass frame.
\par\medskip\noindent
Since in the present experiment the scattered positron and proton
were not detected, the decay angles are determined by approximating the
direction of the virtual photon by that of the
incoming positron. It has been verified by Monte Carlo calculations
that this is a good approximation. 
\par\medskip\noindent
The acceptance corrected
$\phi$ decay angular distributions 
in the kinematic range $0.1 < p^2_T < 0.5 \ \rm{GeV^2}$
are shown in Fig. 6. They have
been fitted with the functions~\cite{angle}:
\[
\frac{1}{N}\cdot\frac{dN}{dcos\theta_{h}} =\frac{3}{4} \left[1-r^{04}_{00}+(3r^{04}_{00}-1)\cos^{2}\theta_{h} \right]
\]   
\[
\frac{1}{N}\cdot\frac{dN}{d\phi_{h}} =\frac{1}{2\pi} (1-2r^{04}_{1-1}\cos2\phi_{h}),
\]
where $r^{04}_{00}$ and $r^{04}_{1-1}$ are 
two of the $\phi$ spin-density matrix
elements.
\par\medskip\noindent
Assuming VDM and $s$-channel helicity conservation, 
the  $r^{04}_{00}$ spin-density matrix
element can be expressed as:
\[
r^{04}_{00} = \frac{Q^2}{M_{\phi}^2} \cdot \frac{\varepsilon}{1+\varepsilon \frac{Q^2}{M_{\phi}^2}},
\]
where $\varepsilon$ is the ratio of the longitudinally to the transversely
polarised photon fluxes.
Assuming the $Q^2$ dependence given in section 2.1,
on average $\varepsilon \simeq 0.998$ and 
$r^{04}_{00} \simeq 0.03$
in the kinematic region under study.
The spin-density matrix element $r^{04}_{1-1}$ is expected to be zero
under the assumption of $s$-channel helicity conservation.
\par\medskip\noindent
The fitted values 
obtained from the distributions in Fig. 6 are
$r^{04}_{00} = -0.01 \pm 0.04$ and 
$r^{04}_{1-1} = 0.03 \pm 0.05$, consistent with VDM and 
$s$-channel helicity conservation.
%
%
\section{Summary and Conclusions}
The photoproduction of $\phi$ mesons has been measured with
the ZEUS detector at HERA.
The cross section is
$\sigma_{\gamma p \rightarrow \phi p} = 0.96 \pm 0.19 ^{+0.21}_{-0.18}$ $\rm{\mu b}$
at $<W>~ = 70 \ \rm{GeV}$ and for $|t| < 0.5$ $\rm{GeV^2}$.
In comparison to lower energy measurements
this result is consistent with Regge theory which predicts
a weak rise of this cross section with increasing $W$ from the
exchange of a soft pomeron.
\par\medskip\noindent
The differential cross section $d\sigma/dt$, determined in the kinematic
range $0.1 < |t| < 0.5$  $\rm{GeV^2}$, falls exponentially
with the slope value $b = 7.3 \pm 1.0 \pm 0.8$  $\rm{GeV^{-2}}$.
The comparison with lower energy data is consistent with the logarithmic
rise of the $t$ slope with $W$ expected by Regge theory.
\par\medskip\noindent
The spin-density matrix elements 
measured from the $\phi$ decay angular distributions
are in argeement
with $s$-channel helicity conservation.
\par\medskip\noindent
At HERA energies, elastic $\phi$ photoproduction shows 
the features typical of a 
soft diffractive reaction. The Regge theory expectations for elastic vector
meson production at HERA energies are thus corroborated at the scale given
by elastic $\phi$ photoproduction.
\section{Acknowledgement}
We thank the DESY Directorate for their strong support and
encouragement. The remarkable achievements of the HERA machine group
were essential for the successful completion of this work and are
gratefully appreciated. 

%
%
\newpage
\begin{figure}
\centerline{
\psfig
{figure=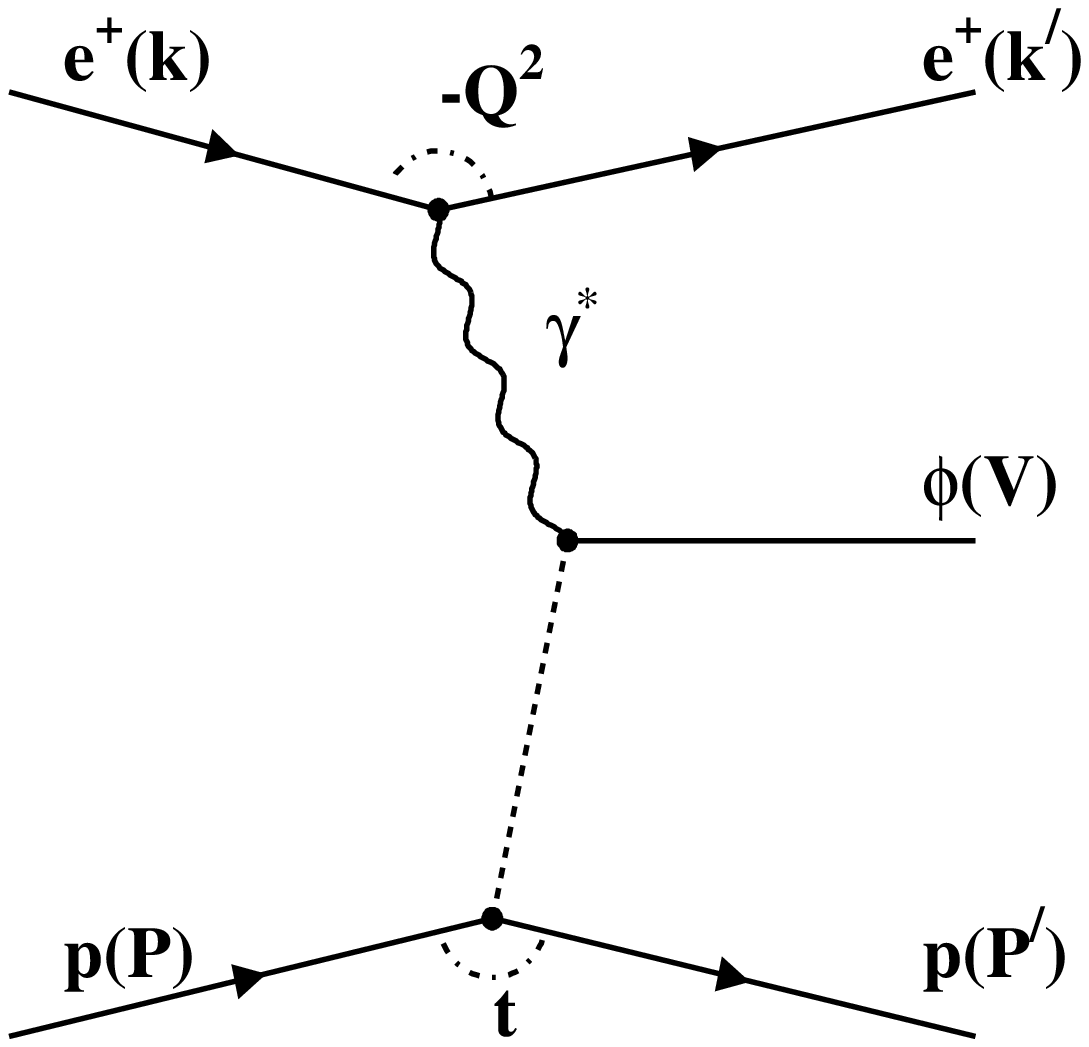,bbllx=1.5cm,bblly=5.0cm,bburx=21.0cm,bbury=25.0cm,height=17.0cm}
}
\caption{Schematic diagram of elastic $\phi$ photoproduction in $e^{+} p$ interactions.}
\end{figure}
%
\newpage
\begin{figure}
\centerline{
\psfig
{figure=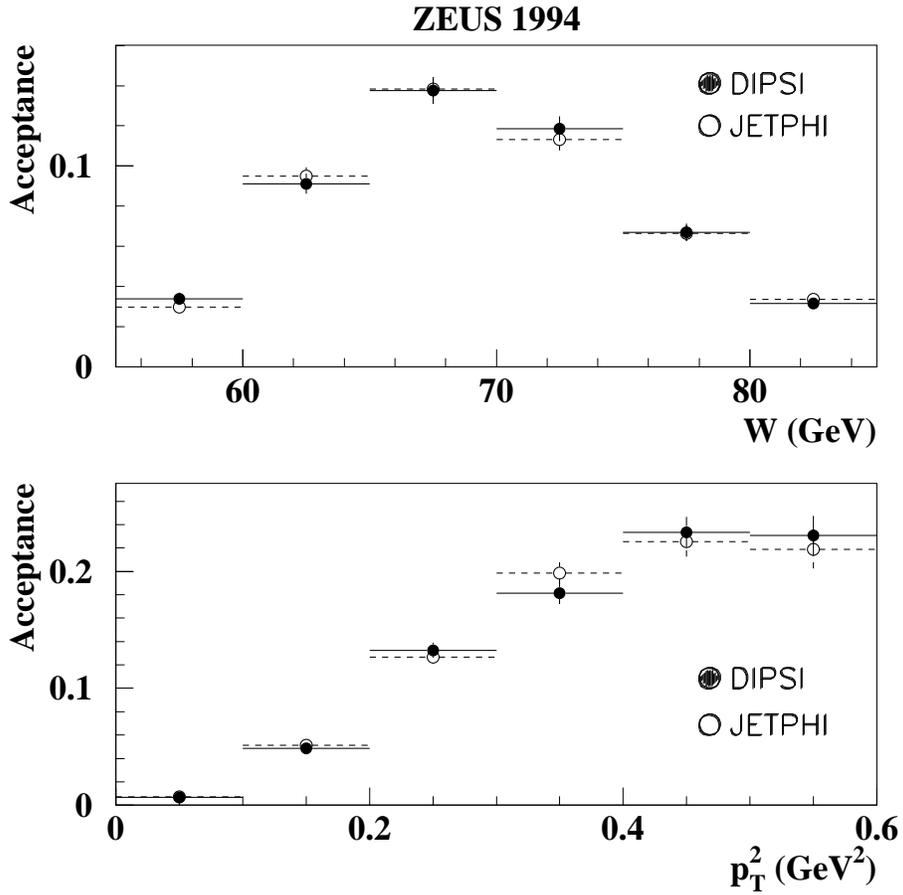,bbllx=1.5cm,bblly=3.0cm,bburx=21.0cm,bbury=21.0cm,height=14.0cm}
}
\caption{Acceptance for the reaction
$e^{+} p \rightarrow e^{+} \phi p$
as a function of $W$ (for $0.1 < p_T^2 < 0.5$ $\rm{GeV^2}$) 
and of $p_{T}^2$ (for $60 < W < 80$ GeV) obtained 
for the two event generators described in the text.}
\end{figure}
%
\newpage
\begin{figure}
\centerline{
\psfig
{figure=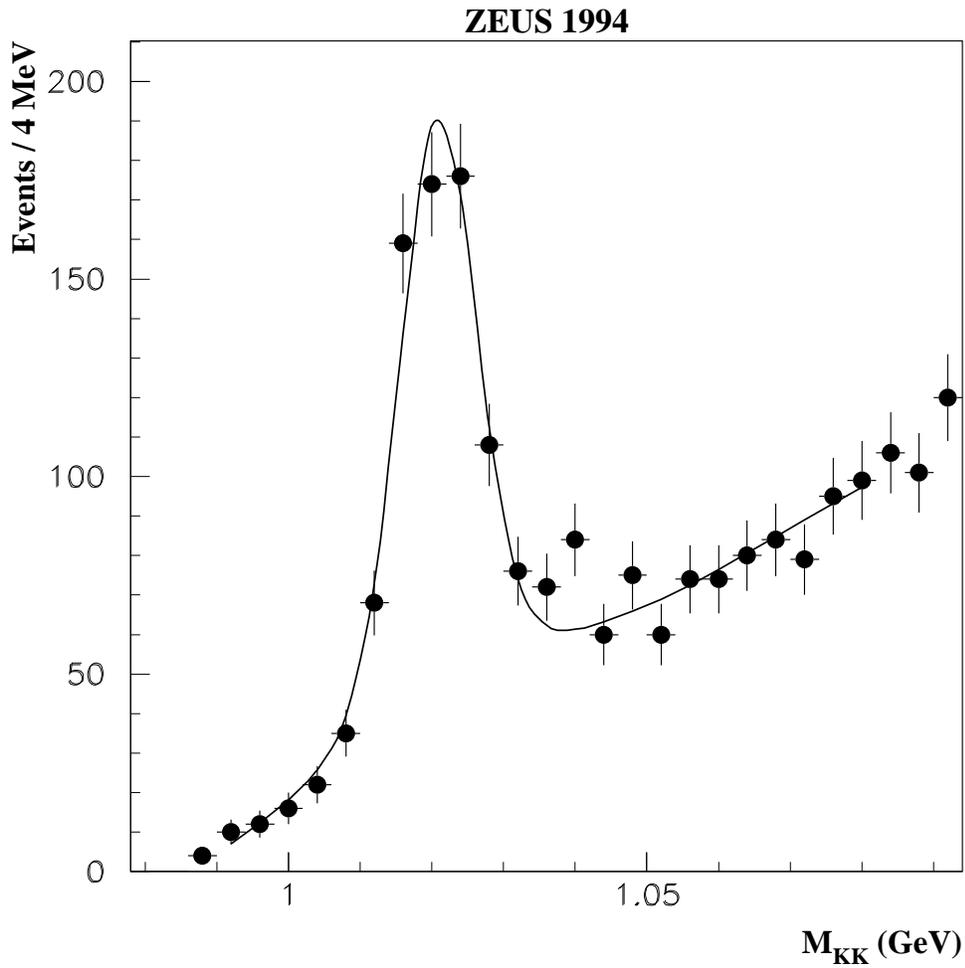,bbllx=1.5cm,bblly=5.0cm,bburx=21.0cm,bbury=25.0cm,height=17.0cm}
}
\caption{Uncorrected $K^+$$K^-$ invariant mass distribution
for all events passing the final selection cuts.
The curve is the result of the fit
described in the text.}
\end{figure}
%
\newpage
\begin{figure}
\centerline{
\psfig
{figure=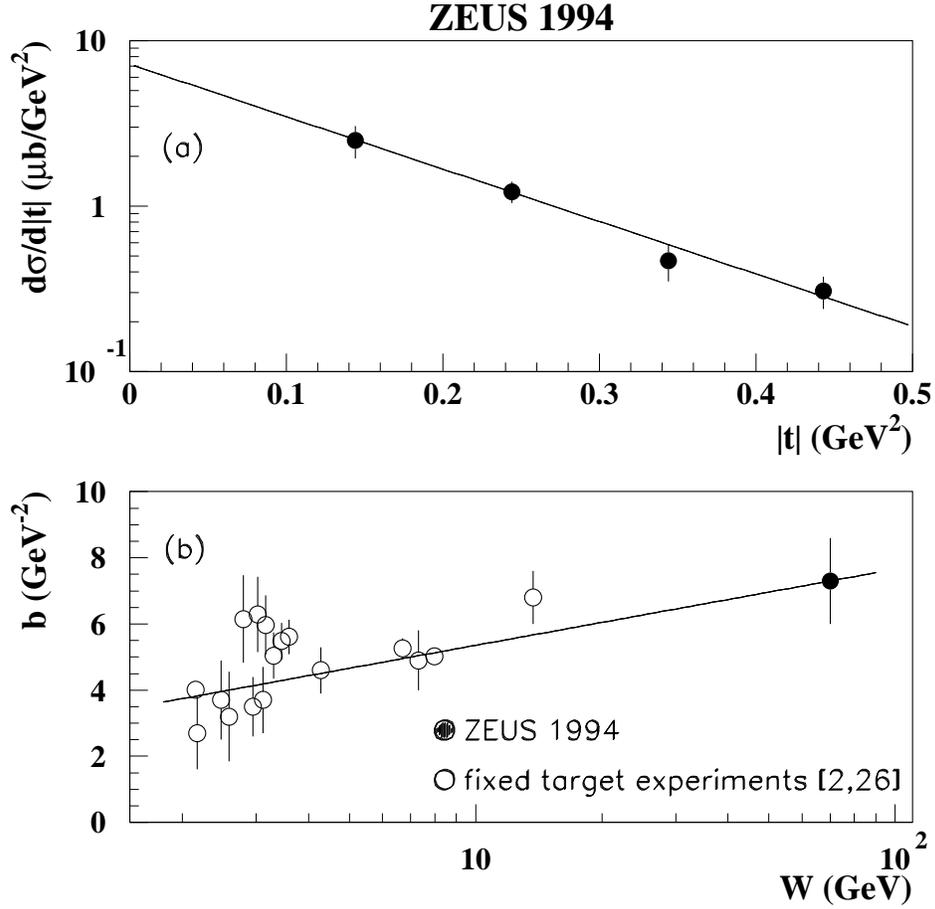,bbllx=1.5cm,bblly=3.0cm,bburx=21.0cm,bbury=23.0cm,height=16.0cm}
}
\vspace*{-1cm}
\caption{(a) Acceptance corrected {\it t} distribution
for the reaction $\gamma p \rightarrow \phi p$ at $<W>~ = 70 \ \rm{GeV}$.
The dots are the ZEUS
data, while the line is the result of the exponential fit described
in the text. (b) Compilation of measurements of the
slope parameter $b$ as a function of $W$ for the reaction
$\gamma p \rightarrow \phi p$. 
The different data are measured in various $t$ intervals.
The line shows the Regge
theory prediction
$b_0 + 4 \alpha' lnW$ with $\alpha' = 0.25$ $\rm{GeV^{-2}}$.
The value for $b_0$ is chosen such that the line intercepts
the ZEUS measurement.}
\end{figure}
%
\newpage
\begin{figure}
\centerline{
\psfig
{figure=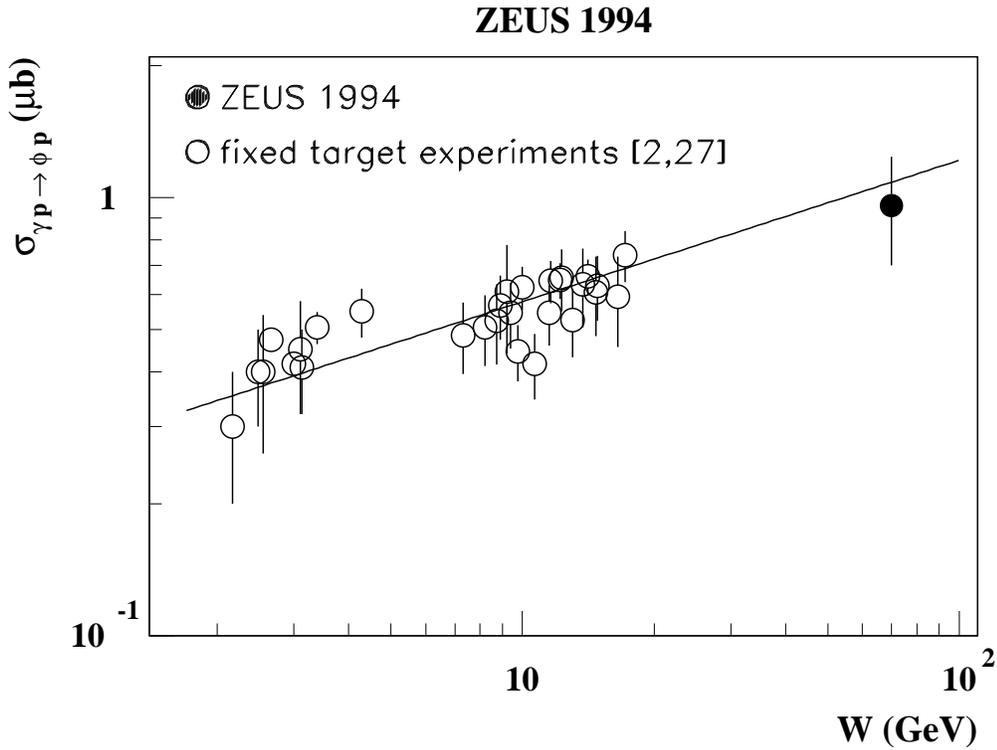,bbllx=1.5cm,bblly=3.0cm,bburx=21.0cm,bbury=23.0cm,height=22.0cm}
}
\vspace*{-9cm}
\caption{Elastic $\phi$ photoproduction cross sections as a function of
$W$. The solid dot is the ZEUS measurement, while the open circles are
the lower energy data. 
The line shown is a description of the fixed target data using
$\sigma_{\gamma p \rightarrow \phi p} \propto W^{0.32}$~\protect\cite{dl}.
It is inspired from Regge theory, which predicts
$\sigma_{\gamma p \rightarrow \phi p} \propto W^{4\epsilon}/b(W)$,
where $1+\epsilon=1.08$ is the intercept of the Regge
trajectory and $b(W)$ is the energy dependent exponential slope of the
differential cross section. This energy dependence however is ignored in the
parametrisation.}
\end{figure}
%
%
\newpage
\begin{figure}
\centerline{
\psfig
{figure=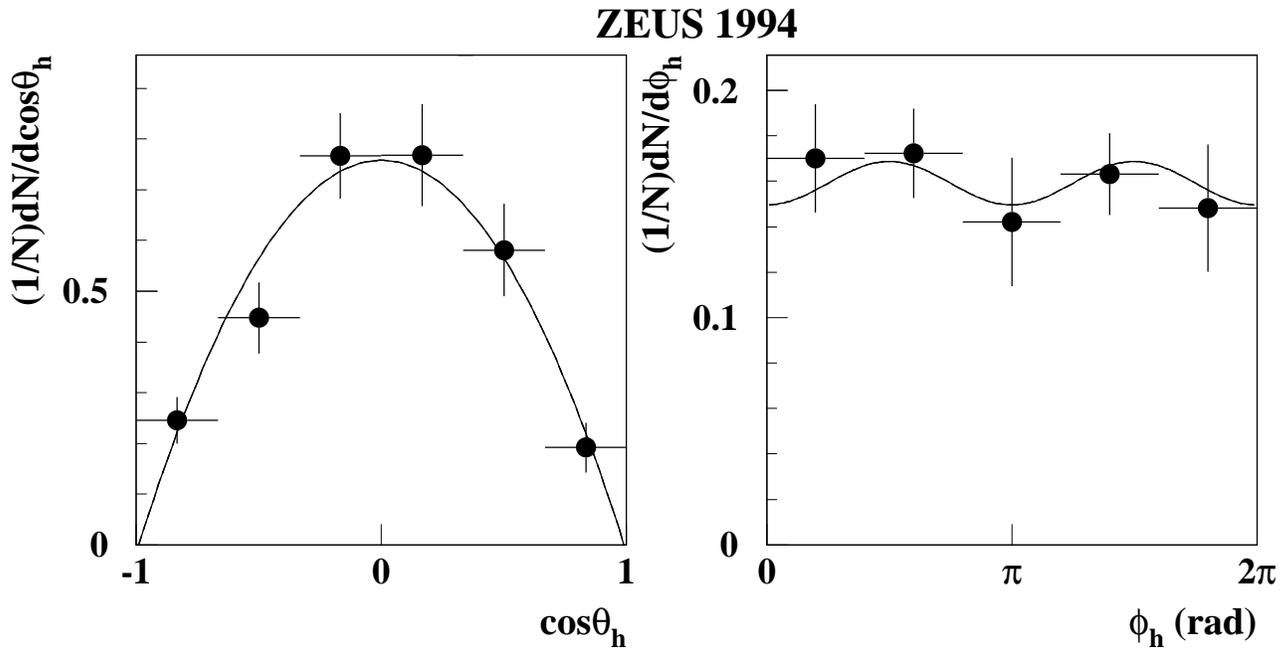,bbllx=1.5cm,bblly=3.0cm,bburx=21.0cm,bbury=23.0cm,height=19.0cm}
}
\vspace*{-7cm}
\caption{Acceptance corrected decay angular distributions
for the $\phi$ meson in the reaction 
$\gamma p \rightarrow \phi p$ at $<W> = 70 \ \rm{GeV}$.
The curves are the results
of the fits described in the text.}
\end{figure}
\end{document}